# STUDENTS' ENGAGEMENT IN AN ANONYMOUS PEER REVIEW: USING THE OPEN-SOURCE SAKAI PLATFORM

Fazlyn Petersen, University of the Western Cape, Fapetersen@uwc.ac.za

Bradley Groenewald, University of the Western Cape, 3658200@myuwc.ac.za

**Abstract:** There is a need to provide quality education without discrimination or prejudice to all students. However, there are challenges in implementing quality education in large classes, especially during remote learning. Literature indicates that providing lecturer feedback can become a tedious task, especially in large classes. Literature states that involving students in the peer review process can improve the quality of their submissions. This research used a case study and thematic analysis. Qualitative data were collected from 179 third-year Information Systems students who used the Opensource Sakai Platform. Students reviewed another student's report, without knowing their identity. The research used self-determination theory as a theoretical basis. The achievement of perceived autonomy is supported as an anonymous peer review helped students to empower themselves. Perceived competence was also achieved as the anonymous peer review improved the quality of work submitted and the development of workplace skills. Perceived relatedness is supported as students indicated that the anonymous peer review allowed them to learn from their peers. It also improved their understanding and the ability to see errors in their work. Despite the negative aspects identified using the Sakai platform, it may provide a viable alternative for providing feedback remotely, especially during the Covid-19 pandemic.

## 1.    INTRODUCTION

The achievement of quality education for all is complex. Authors indicate that the achievement of quality education requires that an equilibrium between the teaching style of the lecturer and the learning style of the student is achieved (Hill, Lomas, & Macgregor, 2003). One such challenge that various institutions of higher learning face are that the number of students who are attending university are and have been growing at an exponential rate (Özoğlu, Gür, & Gümüş, 2016) with a growing level of inequality among student groups.

Students are demographically diverse and digitally literate (Singh, 2016). These are students who are different in terms of race, culture, age, gender, religion, social class, and cognitive domain (Bowman, 2016). Teaching these demographically diverse students is challenging because each student may require different teaching and learning methods in terms of 1. How students assimilate concepts. 2. How students interact with their peers and lecturers. 3. How students and lecturers recognise the differences in terms of demographics (Sadowski, Stewart, & Pediaditis, 2018). To effectively teach diverse, digitally literate students, the literature indicates the importance of providing feedback to provide improved learning in higher education (Jonsson, 2013).

Additionally, lecturers are working remotely, via electronic learning management systems (eLMS), during the Covid-19 pandemic. Therefore, there is an increased need to provide quality feedback to students online.

According to (Baker, 2016), lecturer feedback can become a tedious task as most of the time is spent on fixing grammatical errors. Whereas a minuscule amount of time is spent on addressing students ability to display their understanding of the concepts taught. Also, students do not perceive review in any form as positive, regardless of how constructive the criticism is (Jonsson, 2013). Therefore, if students do not engage with the feedback provided it will not lead to improved student learning and performance (Winterscheid, 2016). An alternative method of providing feedback is the use of peer review.

Peer review is a process to evaluate the work of others to ensure that is it of excellent quality (Harland, Wald, & Randhawa, 2017). Anonymous peer review is another type of peer review process that can be used. The key difference lies in the anonymity of the process (Panadero & Alqassab, 2019). Three benefits can be realised as a result of using an anonymous peer review:
1. It allows students to develop analytical skills,
2. It increases the engagement from the student which could be beneficial in improving the quality of education and
3. If utilised correctly, it enhances learning and the quality of education (Harland et al., 2017).

The literature provides evidence that anonymous peer review increases how students engage with each other and the research (Watkins & Ball, 2018). Anonymous peer review fosters a collaborative and communicative environment that improves learning according to experiential learning theory (Kolb, 2000; Sridharan, Muttakin, & Mihret, 2018). The literature supports that anonymous peer review improves learning and the quality of education. Another study that supports this found that the challenges of anonymous peer review lie in "motivating students to complete the reviews" (Søndergaard & Mulder, 2012). Research has found that 87% of students believe that anonymous peer review is beneficial (Simpson & Clifton, 2016). However, this research was not conducted in a developing country.

South Africa is a developing country where the need to transform higher education has been highlighted (Department of Education, 1996). Due to the legacies of apartheid, there is a historical "inequitable distribution of access and opportunity for students and staff along lines of race, gender, class and geography" (Department of Education, 1997:8). Through the Extension of University Education Act of 1959 new universities were created for students of colour. Students of colour were only permitted entry to white universities where programmes were not offered at black universities (Department of Education, 1996). The inequalities between historically white and black universities are evidenced by the difference in participation rates among different populations, lower ratios of black and female staff and disparities in terms of capacities and facilities (Department of Education, 1997:8). Therefore, "despite the absence of data, it is not unreasonable to think that graduates from historically black universities may struggle to find work more than graduates from universities that served white students because of differences in the perceived quality of their degrees" (Altbeker & Storme, 2013:15).

Therefore, the research question was: *How do students engage in an anonymous peer review in a large class at a historically disadvantaged institution?* The objectives of this research were to determine students' engagement in an anonymous peer review in a large class at a historically disadvantaged institution and to provide recommendations to improve future implementations.

## 2.   RESEARCH DESIGN AND METHODOLOGY

The research utilised a case study design with a qualitative methodology to effectively answer the research question and its stated objectives (Yin, 2003). The aim of this was to understand students as actors and how they engaged in an anonymous peer review. This allowed for "thick descriptions" (Avenier & Thomas, 2015) of students engagement in anonymous peer review to be developed.

Data was gathered from students' textual responses via an online survey from 179 third-year Information Systems students at a historically disadvantaged institution. Students used the Sakai platform to complete the anonymous peer review. The allocation of reports to students was automatically done via the Sakai platform. Students were instructed to remove any identifying information (e.g. names or student numbers) from their submissions. Students did not know the identity of the student whose report they were marking. Students' identities were only known to the lecturer.

The researchers had a low level of control over the students being studied to understand students' engagement with little to no influence. The data analysis used thematic-content analysis which allows student data to be "identified, analysed, organised, described and reported" (Nowell, Norris, White, & Moules, 2017, p. 2).

Ethical considerations were based on the guidelines provided by (Dearden & Kleine, 2018). Students provided consent for participation in this research. The objectives of the study along with the anticipated benefits were explained to all students involved in this study. Student anonymity was maintained by removing unique identifiers, such as student numbers. The collected data was stored in an access restricted folder.

## 3. CASE STUDY: USING THE ANONYMOUS PEER REVIEW ON THE SAKAI PLATFORM

The University of the Western Cape (UWC) is a historically disadvantaged institution that was created for predominantly 'coloured' students. Coloured was a racial classification, based on apartheid. The population consisted of "class of African and Asian origin variously referred to as half-castes, bastards, Cape Boys, off-whites or coloureds" (Adhikari, 2009:xi). However, it also consisted of 'sub-groups such as Malays, Griquas and 'Hottentots' (Adhikari, 2009:xi) who would typically not be allowed to attend the historically white institution, the University of Cape Town. The vice-chancellor in 1987, Jakes Gerwel, referred to UWC as the "*intellectual home for the left*" (Soudien, 2012).

UWC uses the open-source Sakai platform to host their eLMS. The Sakai platform is a free, educational software platform that supports teaching and collaboration (Sakai, 2020). The Sakai platform is regarded as an eLMS leader for three consecutive years (Sakai, 2020). The ability to complete anonymous peer reviews remotely can be accomplished by using platforms such as Turnitin and Sakai. The Sakai platform allows lecturers to use an anonymous peer review as an assessment option (Sakai, 2020). However, if students do not engage with the process having any platform available will not automatically result in the successful completion of a peer review or improving the quality of their work.

Authors provide the following guidelines for the creation of effective, high-quality feedback in Table 1:

**Table 1 Guidelines for the creation of effective, high-quality feedback**

| Guideline | Application in the large, undergraduate course |
|---|---|
| Should be task-related. | An assignment was created, as indicated in the course outline, where the student had to create a report to obtain funding from a potential investor for their business. The anonymous peer review was set as the assessment option. |
| Focused on the quality of the student performance rather than the students' characteristics. | The use of an anonymous peer review reduces the ability of markers to focus on students' characteristics and rather focus on their performance. |
| Must help students to improve their performance. | There was an opportunity provided to complete an optional peer review with fellow class members before the submission of the final assignment on the Sakai platform. If students used this opportunity, they could improve their final submission. A rubric was provided to students to clearly define the marking criteria. |
| Should be delivered in a specific, timely and individualised manner. | The Sakai platform allows the assignment to be released to students at a specific date and time, as indicated in the course outline. The deadline date was set on the platform and included in the course calendar. |
| Feedback will only assist in improving students' performance if it is used. | Students who did not engage in the optional peer review before their final submission performance may not have performed as well as students who completed the exercise. |
| Critical feedback improves the chances of improved student performance (Cutumisu & Schwartz, 2018). | Students received marks and comments to assist them with future submissions. |

## 4. RESEARCH MODEL

A theoretical model is a lens or blueprint which guides the researcher in answering the research question of how students engage in an anonymous peer review and the research objectives (Grant & Osanloo, 2014). It is based on theories from previous studies to guide the development of results in the data analysis (Lederman & Lederman, 2015).

Self-determination theory (SDT) has been used as a model to investigate students' engagement (Reeve, 2012). SDT is constructed from five theories of motivation (Ryan & Deci, 2019):

1. Basic needs theory emphasises the importance of intrinsic motivation, engagement and psychological well-being for achieving psychological needs,
2. Organismic integration theory introduces extrinsic motivation. Extrinsic motivation is the "activities aimed at achieving outcomes separable from the behavior itself" (p.15). While intrinsic motivation refers to the performance of an activity for one's satisfaction,
3. Goal contents theory differentiates between intrinsic goals and extrinsic goals. Intrinsic goals support psychological needs and well-being. Extrinsic goals neglect these needs and results in ill-being,
4. Cognitive evaluation theory predicts the effect of external events on intrinsic motivation and
5. Causality orientations theory identifies how individual differences influences motivation.

Based on SDT, the research model focused on three areas, namely:

1. Autonomy refers to people's need to feel in control of their behaviour and goals,
2. Competence is the need to master tasks and to learn different skills and
3. Relatedness refers to the need to feel a sense of belonging and a connection to other people (Deci & Ryan, 2008).

The application of SDT to the research was completed as displayed in the research model (refer to Figure 1):

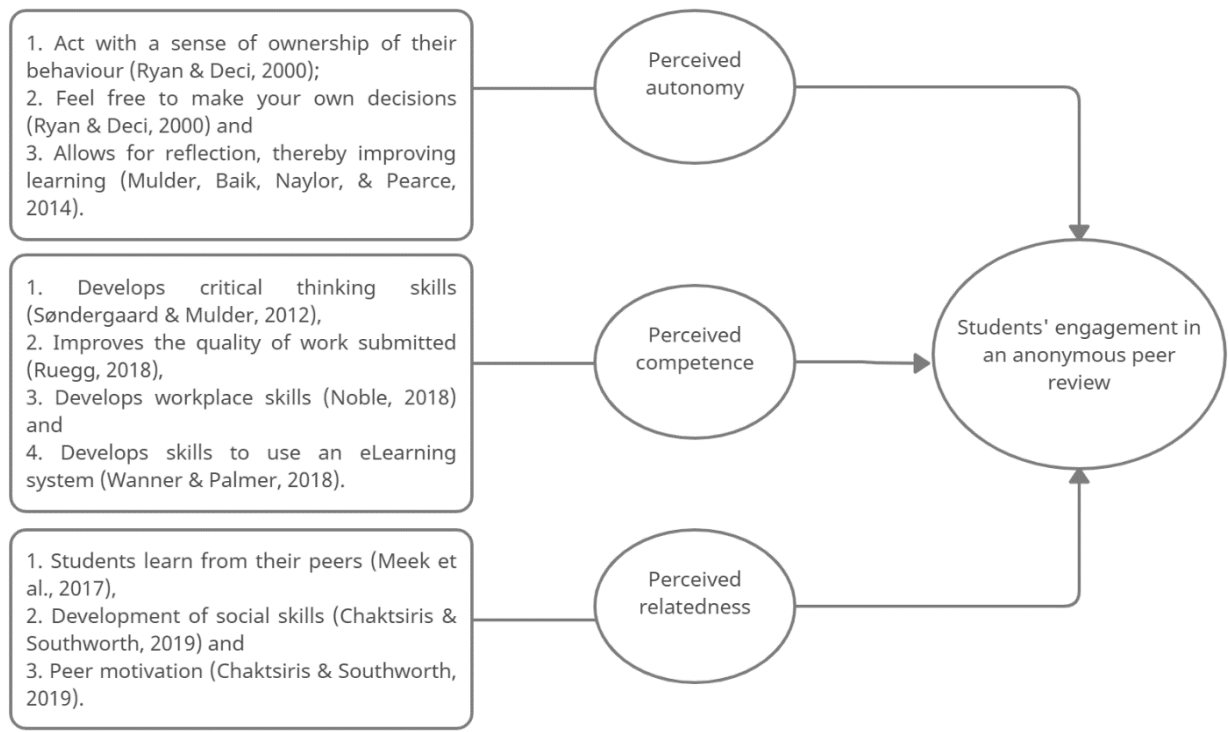

**Figure 1 Research model**

## 5. RESULTS

An analysis of demographic data for 179 third-year B.Com Information Systems students who participated in the survey was completed. Results indicated that the majority of students are between 16-24 years old (80.4%) and the minority between 35-49 years old (4.5%). The majority of students are male (54.7%) and full-time students (87.7%).

### 5.1. Perceived autonomy

Based on the research model, students' level of perceived autonomy is based on three factors, as specified in the research model.

#### 5.1.1. Acts with a sense of ownership of their behaviour

Evidence indicates that the anonymous peer review helps students to empower themselves. This is supported by students' quotes: *"I actually felt empowered reviewing another student's work"* (female, 16-24) and *"Peer review might sound unnecessary but another students review would make you feel proud about your own work or make you want to work even harder because that student's work is inspiring"* (female, 16-24).

#### 5.1.2. Feel free to make own decisions

Perceived autonomy is positively supported by students' ability to make their own decisions. Quotations included, *"We all take different decisions and have opinions which differs but at the end we all can reach the destination. It proves that they is not just one formula in life they are other ways to get things done not just one way"* (female, 16-24). However, the data is indicated that students also did not trust the anonymous peer review exercise. This is due to various factors, one of them being anonymous peer review gives students too much decision-making ability in marking assignments. This was highlighted by a student who said the following: *"…it was frustrating to see* *how other people did their assignments and have to think that those same people who are not good writers have to critique your assignment"* (female, 16-24).

Evidence highlighted that there was an element of unfairness in the anonymous peer review exercise. This is summarised in the following quote: *"The peer review exercise was okay, I did not enjoy marking the other persons. I feel like the peer review is not fair to some students, some will mark wrong and others right. It would have been better if the lecturer or the tutors marked our assignments"* (female, 16-24).

### 5.1.3. Anonymous peer review allows for reflection

In the data, it was found that students perceived their learning to improve by being able to compare assignments. This finding is largely supported by the majority as summarised in the following quote: *"I felt that it helped me judge myself and better my mistakes as I read through the other projects"* (male, 16-24). Students also learnt the importance of following instructions, *"It actually showed me how we actually lose marks by not following instructions"* (male, 16-24).

## *5.2.  Perceived competence*

Students' engagement in the anonymous peer review was analysed based on their level of perceived competence. Perceived competence includes four determinants, as indicated in the research model.

The findings of the research highlighted that there was a level of perceived competence achieved for the anonymous peer review. The following section will describe the themes identified in the research model and the application to the data.

### 5.2.1. Develops critical thinking and analysis skills

The data analysed indicates that students perceive anonymous peer review to aid them in developing critical thinking and analysis skills. This is summarised by the majority of the students who are males between the ages of 16-24 in the following quotes: "*The peer review exercise taught me how to read with understanding and pushed me to improve my analytical posture*".

It was also found that students learnt how to be more objective in analysing and reviewing their peers' work. This theme was highlighted by the following quote: *"Forces one to be objective"* (male, 16-24). However, the results also indicated that students did not understand the course content that was provided and thus could not effectively perform the anonymous peer review. Therefore, students did not provide adequate or any feedback during the anonymous peer review.

### 5.2.2. Develops workplace skills

Students indicated that the anonymous peer review aided them in developing workplace skills. The findings highlighted management and time-management skills. In support of this are two quotes from students: *"I gained management skills"* (female, 25-34) and *"Learning how to best manage my time… qualities that will come handy in the workplace"* (male, 25-34). However, it was also found that students are under the impression that not all of their peers can write reports. This negatively affected student engagement with the anonymous peer review. A student noted the following: *"I learnt that not many of my peers actually know how to write a business report"* (male, 16-24). Therefore, the ability of such students to conduct a peer review was questioned.

People skills was another social skill highlighted by a student who stated the following: *"I did this in a respectful manner which built my people skills"* (female, 16-24). In addition to this, it was found that anonymous peer review helps students to understand marking from the perspective of the reviewer.

### 5.2.3. Improves the quality of work submitted

The findings highlighted student perceptions that there was an improvement in the quality of their work. This is summarised by the following quote:
*"I enjoyed it. It helped me to understand what I did wrong in my own report. It also helped me measure my own standards by peer reviewing someone else's report"* (female, 16-24).

The data also indicated that students found the anonymous peer review easy to understand. This can be summarised by a common quote from the majority of students: *"It was quite straightforward and simple to do"* (male, 16-24). Students' engagement with the anonymous peer review decreased in complexity and become easier as they neared completion of the exercise. This is highlighted in the following statement from a student: "*But when I realised how to do it using the rubric I was able to do it easily*" (male, 16-24). However, findings also indicated that students found the anonymous peer review to be difficult and uncomfortable.

Although the majority of students perceived anonymous peer review to improve their learning, it is suggested that some students perceived the anonymous peer review exercise to be confusing and uncertain.

### 5.2.4. Develop skills to use eLearning systems

A finding that negatively impacted student engagement was that the majority of students found that they could not complete the anonymous peer review due to issues related to the Sakai platform used. Students' views were summarised in the following quote: *"It's a bit difficult to do when the system itself made it difficult to complete"* (male, 16-24). Although this issue appeared the data suggests that some students solved it by getting assistance from the lecturer. This is highlighted by a student who said the following: *"This caused frustrations however the lecturer assisted me"* (female, 16-24).

## 5.3. *Perceived relatedness*

The level of perceived relatedness was also investigated to assess students' engagement in the anonymous peer review.

### 5.3.1. Students to learn from their peers

The finding is supported by the majority of students (78 quotations) who indicated that the anonymous peer review exercise allowed them to learn from their peers. This can be summarised by the following quote: *"I got to see how other people tackled the problem. It also exposed me to thinking from different perspectives. Also adding to my learning experience"* (male, 16-24). It was also found that students were able to understand how their peers thought, learnt and had differing ideas about the assignment.

### 5.3.2. Development of social skills

Evidence indicates that there are students who preferred to be marked by their peers. This finding was supported by the following quote, *"It was really good to get marked by the student"* (female, 16-24). However, contrary evidence indicates that students did not trust their peers enough to review each other's work. The data analysed indicated that students themselves feel a level of uncertainty when it comes to anonymous peer review. This is highlighted in a quote from the following student: *"Would prefer not having to do it again. I didn't know if I did it correct[ly]"* (female, 35-49). The lack of trust finding could be improved if students adhered to or understood the instructions.

### 5.3.3. Peer motivation

Perceived relatedness was negatively impacted by students' perceiving that their peers put in less effort than them. This factor is supported by the following quote, *"I strongly disagree with this exercise, the anonymous marker may not be up to speed with the course work and may not know what you are writing about and it can put hardworking students in the deep end"* (male, 16-24).

Despite the anonymous peer review being a compulsory part of the assignment not all students completed it by the deadline or complete it at all. This resulted in students who had completed their anonymous peer review but not receiving their results. It led to having to motivate students to complete the anonymous peer review. Evidence of this can be summarised in the following student comment: *"Next time I will try and ensure that my peers meet the deadline so that it does not affect my work"* (female, 16-24).

## 6. DISCUSSION

The findings indicated that perceived autonomy positively influenced students' engagement with the anonymous peer review. Anonymous peer review also allows for students to reflect on their work is supported by (Wanner & Palmer, 2018).

Evidence indicates that students' engagement was possibly influenced by perceived competence. Students believed that the anonymous peer review improved their learning, aid in the development of skills and improves the quality of work submitted. These findings are supported by (Noble, 2018) who found that the majority of students felt that their written work had improved. The literature states that involving students in the peer review process improves the quality of the result of their submission (Baker, 2016). This is because students are aware of others, their peers reading their work and thus put more effort into it (Watkins & Ball, 2018).

The authors also found that anonymous peer review aids in the development of skills such as critical thinking (Simpson & Clifton, 2016). Findings also supported the development of workplace skills and social skills, as indicated in the literature (Chaktsiris & Southworth, 2019). It should be noted that a certain level of uncertainty was created by when peers mark. This resulted in the anonymous peer review being difficult and uncomfortable. This also supported by (Meek et al., 2017) as they found that peers are not qualified enough to perform the review and that it is uncomfortable reviewing peers. Although, this finding is supported by the minority it should not be dismissed in attempting to improve the anonymous peer review process.

Positive perceived related themes resulted in students improving their learning by learning from peers. This finding corroborates with a finding by (Meek et al., 2017) as they found that anonymous peer review allows students to see a different viewpoint and seeing good examples of writing. However, a negative finding that impacts perceived relatedness are peers not trusting each other. Additionally, not all students were motivated enough to complete the review. This supports the above finding in that students did not trust their peers due and not all students completing the anonymous peer review. The majority of students enjoyed reviewing their peers and viewed it as beneficial.

The last notable finding is that there were issues related to the technology used in the anonymous peer review exercise. Students were required to complete their anonymous peer reviews using the open-source Sakai platform. The authors found that the reliability and validity of the peer review were put into question when the technology used does not work (Søndergaard & Mulder, 2012; Wanner & Palmer, 2018). However, the lecturer worked with Sakai support staff to address the matter and assist students. Using an open-source eLMS may be a viable option for other historically disadvantaged institutions, due to the lower costs such as not paying for licences. At UWC, the Sakai platform was zero-rated to allow students access to remote learning during the pandemic.

# 7. CONCLUSIONS

The benefits of the peer review were discussed as a means of achieving quality education. Ultimately, it was discovered that there is not sufficient literature on this topic on the implementation of anonymous peer review in large classes in a historically disadvantaged institution. Therefore, this research adds to the body of knowledge by examining students' engagement in an anonymous peer review. The research findings can be used to identify areas for future improvement.

This research aimed at answering the research question by identifying students' engagement with an anonymous peer review exercise using the open-source Sakai platform. The research used a case study research design and qualitative methodology.

The findings in the research highlighted that the anonymous peer review helped students develop critical skills related to thinking, working and socialising. It was also found that it improves students' learning through reflection and students learning from each other. Although there were positive findings, negative findings also emerged. These findings are that students do not trust the anonymous peer review and that they are uncertain about its use. Also, it was found that the Sakai platform used in the anonymous peer review could lead to further challenges.

The findings are limited to third-year students at a historically disadvantaged institution and therefore cannot be generalised. Recommendations for future research are that a mixed-methods approach should be used. This is due to the effectiveness of a mixed-methods approach in effectively understanding the data. Moreover, future studies should gather data from postgraduate students as well. This will allow researchers to evaluate the efficacy of using anonymous peer review in both undergraduate and post-graduate courses.

*Petersen & Groenewald*     *Students' Engagement in an Anonymous Peer Review*
Department of Education. (1996). Green Paper on Higher Education Transformation. Retrieved from http://scholar.google.com/scholar?hl=en&btnG=Search&q=intitle:GREEN+PAPER+ON+HIGHER+EDUCATION+TRANSFORMATION#3

Department of Education. (1997). White paper 3: A programme for higher education transformation. Government Gazette (Vol. 58). Pretoria. https://doi.org/10.1038/058324a0

Grant, C., & Osanloo, A. (2014). Understanding, Selecting, and Integrating a Theoretical Framework in Dissertation Research: Creating the Blueprint for Your "House." Administrative Issues Journal Education Practice and Research, 12–26. https://doi.org/10.5929/2014.4.2.9

Harland, T., Wald, N., & Randhawa, H. (2017). Student peer review: enhancing formative feedback with a rebuttal. Assessment and Evaluation in Higher Education, 42(5), 801–811. https://doi.org/10.1080/02602938.2016.1194368

Hill, Y., Lomas, L., & Macgregor, J. (2003). Students' perceptions of quality in higher education. Quality Assurance in Education, 11(1), 15–20. https://doi.org/10.1108/09684880310462047

Jonsson, A. (2013). Facilitating productive use of feedback in higher education. Active Learning in Higher Education, 14(1), 63–76. https://doi.org/10.1177/1469787412467125

Kolb, D. A. (2000). The Process of Experiential Learning. In Strategic Learning in a Knowledge Economy (pp. 313–331). https://doi.org/10.1016/b978-0-7506-7223-8.50017-4

Lederman, N. G., & Lederman, J. S. (2015). What Is A Theoretical Framework? A Practical Answer. Journal of Science Teacher Education, 26(7), 593–597. https://doi.org/10.1007/s10972-015-9443-2

Meek et al. (2017). Is peer review an appropriate form of assessment in a MOOC? Student participation and performance in formative peer review. Assessment and Evaluation in Higher Education, 42(6), 1000–1013. https://doi.org/10.1080/02602938.2016.1221052

Noble, A. (2018). Formative Peer Review: Promoting Interactive, Reflective Learning or the Blind Leading the Blind? University of Detroit Mercy Law Review, 94(3), 440–454.

Nowell, L. S., Norris, J. M., White, D. E., & Moules, N. J. (2017). Thematic Analysis: Striving to Meet the Trustworthiness Criteria. International Journal of Qualitative Methods, 16(1), 1–13. https://doi.org/10.1177/1609406917733847

Özoğlu, M., Gür, B. S., & Gümüş, S. (2016). Rapid Expansion of Higher Education in Turkey: The Challenges of Recently Established Public Universities (2006–2013). Higher Education Policy, 29(1), 21–39. https://doi.org/10.1057/hep.2015.7

Panadero, E., & Alqassab, M. (2019). An empirical review of anonymity effects in peer assessment, peer feedback, peer review, peer evaluation and peer grading. Assessment and Evaluation in Higher Education, 0(0), 1–26. https://doi.org/10.1080/02602938.2019.1600186

Ryan, R. M., & Deci, E. L. (2019). Brick by Brick: The Origins, Development, and Future of Self-Determination Theory. In Advances in Motivation Science (pp. 111–156). https://doi.org/10.1016/bs.adms.2019.01.001

Sadowski, C., Stewart, M., & Pediaditis, M. (2018). Pathway to success: using students' insights and perspectives to improve retention and success for university students from low socioeconomic (LSE) backgrounds. International Journal of Inclusive Education, 22(2), 158–175. https://doi.org/10.1080/13603116.2017.1362048

Sakai. (2020). Sakai Learning Management System. Retrieved March 24, 2021, from https://www.sakailms.org/

Simpson, G., & Clifton, J. (2016). Assessing postgraduate student perceptions and measures of learning in a peer review feedback process. Assessment and Evaluation in Higher Education, 41(4), 501–514. https://doi.org/10.1080/02602938.2015.1026874

Singh, R. J. (2016). Current trends and challenges in South African higher education. South African Journal of Higher Education, 29, 1–7. https://doi.org/10.20853/29-3-494
Proceedings of the 1st Virtual Conference on Implications of Information and Digital Technologies for Development, 2021

175